\documentclass[aps,prd,onecolumn,superscriptaddress,nofootinbib,showpacs]{revtex4}

\usepackage{amsmath,bbm,latexsym,amssymb}
\usepackage{graphicx}

\newcommand{\be}{\begin{equation}}
\newcommand{\ee}{\end{equation}}
\newcommand{\ba}{\begin{eqnarray}}
\newcommand{\ea}{\end{eqnarray}}

\newcommand{\ra}{\rightarrow}

\begin{document}
\vspace*{1cm}

\title{Constraining wrong-sign $hbb$ couplings with $h \rightarrow \Upsilon \gamma$}

\author{Tanmoy Modak}\thanks{E-mail: tanmoyy@imsc.res.in}
\affiliation{The Institute of Mathematical Sciences,
Chennai, India}
\author{Jorge C.\ Rom\~{a}o}\thanks{E-mail: jorge.romao@tecnico.ulisboa.pt}
\affiliation{CFTP, Departamento de F\'{\i}sica,
Instituto Superior T\'{e}cnico, Universidade de Lisboa,
Avenida Rovisco Pais 1, 1049 Lisboa, Portugal}
\author{Soumya Sadhukhan}\thanks{E-mail: soumyasad@imsc.res.in}
\affiliation{The Institute of Mathematical Sciences, Chennai, India}
\author{Jo\~{a}o P.\ Silva}\thanks{E-mail: jpsilva@cftp.ist.utl.pt}
\affiliation{CFTP, Departamento de F\'{\i}sica,
Instituto Superior T\'{e}cnico, Universidade de Lisboa,
Avenida Rovisco Pais 1, 1049 Lisboa, Portugal}
\author{Rahul Srivastava}\thanks{Email: rahuls@prl.res.in}
\affiliation{The Institute of Mathematical Sciences, Chennai, India}
\affiliation{Physical Research Laboratory, Ahmedabad, India }

\date{\today}

\begin{abstract}
The rare decay $h \rightarrow \Upsilon \gamma$ has a
very small rate in the Standard Model,
due to a strong cancellation between the direct and indirect
diagrams.
Models with a changed $hbb$ coupling can thus lead to a great increase
in this decay.
Current limits on two Higgs doublet models still allow for the possibility that
the $hbb$ coupling might have a sign opposite to the Standard Model;
the so-called ``wrong-sign''.
We show how $h \rightarrow \Upsilon \gamma$ can be used to put
limits on the wrong-sign solutions.
\end{abstract}


\maketitle

\section{\label{sec:intro}Introduction}

With the discovery at LHC of the first spin 0 particle
\cite{ATLASHiggs, CMSHiggs},
one must now probe its couplings in detail,
searching for discrepancies with the
Standard Model (SM) Higgs.
Of particular interest is the possibility that
the $hbb$ coupling could have a magnitude close to the SM value,
but opposite sign; the ``wrong-sign'' solution.
Current data is consistent with this possibility
\cite{Espinosa:2012im, Falkowski:2013dza,
Belanger:2013xza}.

There is great interest in the two Higgs doublet model
(2HDM) \cite{hhg, ourreview}.
Most attention is devoted to models with
a discrete $Z_2$ symmetry, softly broken by a term
with a real coefficient.
These models have two charged scalars $H^\pm$,
one pseudoscalar $A$,
a heavy scalar $H$,
and a light scalar $h$,
which we identify as the 125 GeV scalar from LHC.
There are four types of such models. Of these,
only Type II and Flipped are consistent with
wrong-sign solutions
\cite{Celis:2013ixa, Ferreira:2014naa, Ferreira:2014dya}.

Naturally,
a sign change does not affect the $h \rightarrow b \bar{b}$
rate,
which,
in most models of the 125 GeV scalar,
is very close to its total width.
Thus,
the effect of the wrong-sign must be sought
indirectly,
for example through its one-loop contribution
to the glue-glue production $g g \rightarrow h$
and di-photon decay $h \rightarrow \gamma \gamma$.
However,
there, loops with intermediate bottom quarks
compete with much larger contributions from
loops with top quarks ($g g \rightarrow h$) or
with top quarks and with gauge bosons ($h \rightarrow \gamma \gamma$).
As a result,
these processes will have values close to the SM,
and only a very precise measurement of order 5\% in
$p p \rightarrow h \rightarrow \gamma\gamma$ will enable experiments
to disentangle the normal sign from the wrong-sign
solutions \cite{Ferreira:2014naa, Fontes:2014tga}.

In contrast, the rare decay $h \rightarrow \Upsilon \gamma$
involves two diagrams which have almost the same magnitude in the SM.
The decay is very suppressed in the SM
(compared, for example, with $h \rightarrow J/\psi\, \gamma$)
due to an accidental cancellation between the two diagrams
\cite{Bodwin:2013gca, Bodwin:2014bpa}.
A change in the $hbb$ sign will destroy the precise cancellation and
will have a dramatic effect in this decay,
making $h \rightarrow \Upsilon \gamma$ the prime candidate to probe the
wrong-sign solutions.
The importance of such a measurement on the wrong-sign solutions
of the 2HDM is the subject of this article.

In Section~\ref{sec:notation} we introduce
our notation,
and in Section~\ref{sec:decay} we present the details
of the $h \rightarrow \Upsilon \gamma$ decay and
perform a full simulation within the real 2HDM.
In Section~\ref{sec:conclusions} we draw our conclusions.

\section{\label{sec:notation}Wrong-sign solution in the 2HDM}

\subsection{Notation}

In this article we consider a CP-conserving
2HDM with a discrete $Z_2$ symmetry,
broken softly by a real term, reviewed extensively for example
in \cite{hhg,ourreview}.
The scalar potential may be written as
\ba
V_H
&=&
m_{11}^2 |\Phi_1|^2
+ m_{22}^2 |\Phi_2|^2
- m_{12}^2\, \left[\Phi_1^\dagger \Phi_2 + \Phi_2^\dagger \Phi_1\right]
\nonumber\\*[2mm]
&&
+\, \frac{\lambda_1}{2} |\Phi_1|^4
+ \frac{\lambda_2}{2} |\Phi_2|^4
+ \lambda_3 |\Phi_1|^2 |\Phi_2|^2
+ \lambda_4\, (\Phi_1^\dagger \Phi_2)\, (\Phi_2^\dagger \Phi_1)
\nonumber\\*[2mm]
&&
+\,
\frac{\lambda_5}{2} \left[(\Phi_1^\dagger \Phi_2)^2 +
(\Phi_2^\dagger \Phi_1)^2\right],
\label{VH}
\ea
with all coefficients real.
The vacuum expectation values (vevs) are also real and written as
$v_1/\sqrt{2}$ and $v_2/\sqrt{2}$.
The fields may be parametrized in terms of the mass eigenstates
as
\ba
\Phi_1 &=&
\left(
\begin{array}{c}
c_\beta G^+ - s_\beta H^+\\*[2mm]
\tfrac{1}{\sqrt{2}}\left[ v c_\beta +
\left( - s_\alpha h + c_\alpha H\right)
+ i \left( c_\beta G^0 - s_\beta A\right)
\right]
\end{array}
\right),
\nonumber\\
\Phi_2 &=&
\left(
\begin{array}{c}
s_\beta G^+ + c_\beta H^+\\*[2mm]
\tfrac{1}{\sqrt{2}}\left[ v s_\beta +
\left( c_\alpha h + s_\alpha H\right)
+ i \left( s_\beta G^0 + c_\beta A\right)
\right]
\end{array}
\right),
\ea
where $c_\theta$ ($s_\theta$) is the cosine (sine) of any angle
$\theta$ in subscript,
$\tan{\beta}=v_2/v_1$,
and $v = \sqrt{v_1^2 + v_2^2} = (\sqrt{2} G_F)^{-1/2}$.
The fields $G^\pm$ and $G^0$ are the would-be Goldstone bosons.

We assume that the lightest scalar ($h$) is the 125 GeV resonance found
at LHC. Its couplings with the gauge bosons are
\be
{\cal L}_{hVV} =
\sin{(\beta - \alpha)} h
\left[
\frac{m_Z^2}{v} Z^\mu Z_\mu
+ 2\, \frac{m_W^2}{v} W^{+ \mu} W^-_\mu
\right].
\ee
The SM limit corresponds to $\sin{(\beta - \alpha)}=1$.
We are interested in models with wrong-sign solutions for the fermion couplings.
Given current experiments,
only Type II and Flipped are consistent with this possibility
\cite{Celis:2013ixa, Ferreira:2014naa, Ferreira:2014dya}.
In these models,
the couplings of $h$ with the fermions from the
third family are
\be
-{\cal L}_\textrm{Yuk}
=
\frac{m_t}{v} k_U\, h \bar{t} t
+
\frac{m_b}{v} k_D\, h \bar{b} b
+
\frac{m_\tau}{v}\, k_\tau\, h \tau^+ \tau^-\, ,
\ee
where
\be
k_U = \frac{\cos{\alpha}}{\sin{\beta}},\ \ \
k_D = - \frac{\sin{\alpha}}{\cos{\beta}}.
\label{kUkD}
\ee
The only difference between the Type II and Flipped models
lies in the coupling of the charged fermions, given,
respectively, by
\be
k_\tau = k_D\ (\textrm{Type II})\, ,
\ \ \ \
k_\tau = k_U\ (\textrm{Flipped})\, .
\ee
The SM limit is $k_U=k_D=k_\tau=1$.

We will denote the ratios between
the 2HDM and SM rates by
\be
\mu_f
=
\frac{\sigma^\textrm{2HDM}(pp \ra h)}{\sigma^\textrm{SM}(pp \ra h)}
\frac{\Gamma^\textrm{2HDM}[h \ra f]}{\Gamma^\textrm{SM}[h \ra f]}
\frac{\Gamma^\textrm{SM}[h \ra \textrm{all}]}{
\Gamma^\textrm{2HDM}[h \ra \textrm{all}]}
,
\label{mus}
\ee
where $\sigma$ is the cross section for Higgs production,
$\Gamma[h \ra f]$ the decay width into the final state $f$,
and $\Gamma[h \ra \textrm{all}]$ is the total Higgs decay width.

\subsection{\label{subse:muVV}A naive explanation for the wrong-sign}

For simplicity, let us assume that the production of $h$ is due exclusively
to the gluon fusion process with intermediate top quark,
and that its width is due exclusively to the decay $h \rightarrow b \bar{b}$.
Within these assumptions
\be
\sqrt{\mu_{VV}} = \pm \frac{k_U}{k_D} \sin{(\beta-\alpha)},
\label{factors}
\ee
where the sign (which will be ignored henceforth) is chosen to make
the square root positive.
Imagine that $\mu_{VV} \sim 1$ because both factors are close to unity.
We start by noting that
\be
-\frac{k_U}{k_D} = \frac{1}{t_\alpha t_\beta}
=
\frac{\cos{(\beta - \alpha)} + \cos{(\beta + \alpha)}}{
\cos{(\beta - \alpha)} - \cos{(\beta + \alpha)}},
\ee
where $t_\theta$ is the tangent of the angle $\theta$.
We find that $|k_U/k_D| \sim 1$ if $\beta - \alpha = \pi/2$,
in which case $k_D=+1$ (the right-sign solution),
or else if $\beta + \alpha = \pi/2$,
in which case $k_D=-1$ (the wrong-sign solution).

\begin{figure}[h!]
	\centering
	\includegraphics[width=0.4\linewidth]{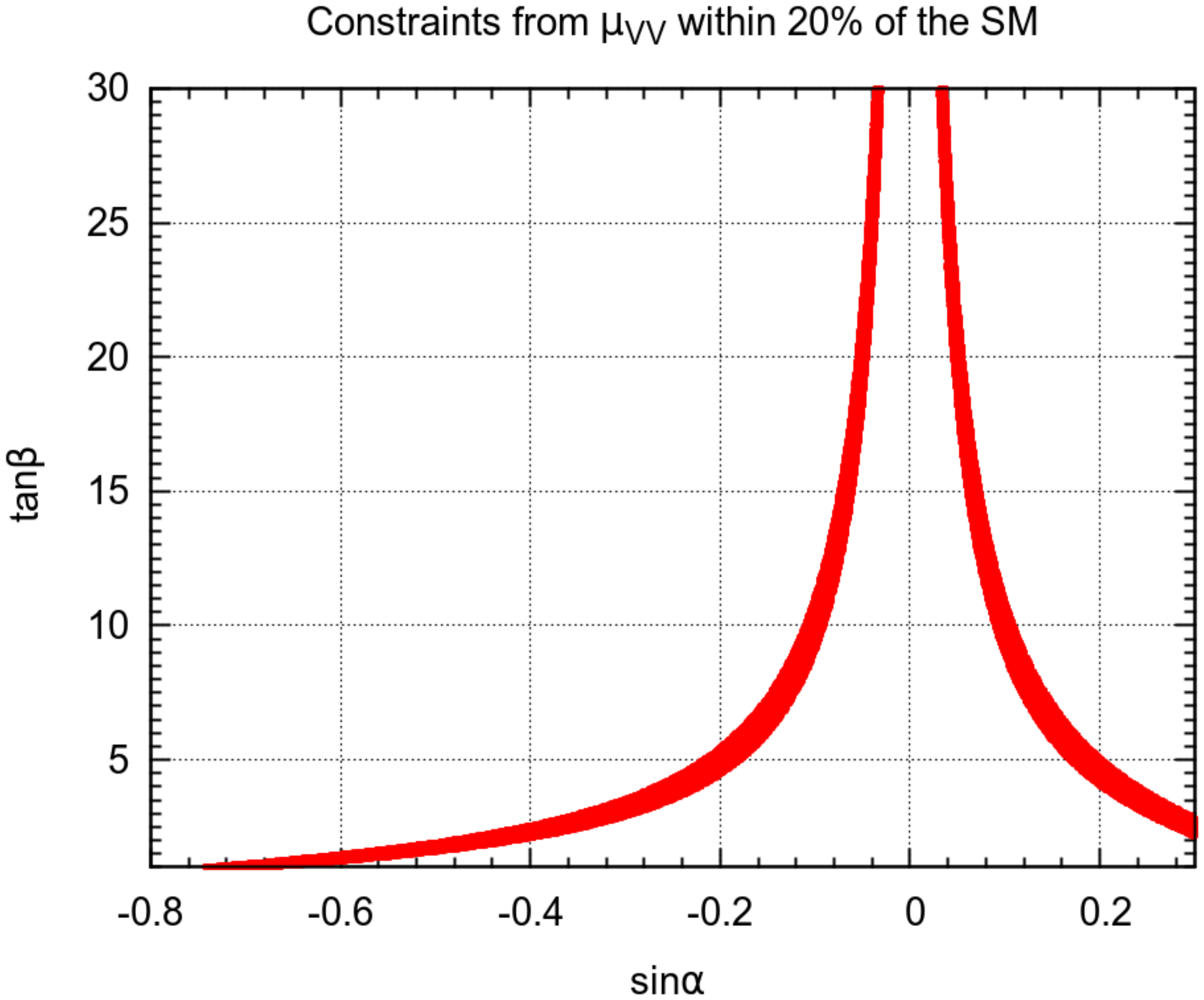}
	\caption{Constraints from $0.8 \leq \mu_{VV} \leq 1.2$ on the
    $\sin{\alpha}$-$\tan{\beta}$ plane.}
	\label{fig:1}
\end{figure}

Now, we look at the second factor in Eq.~\eqref{factors}.
We find
\be
\frac{\sin{(\beta - \alpha)}}{\sin{(\beta + \alpha)}}
= \frac{1-\frac{t_\alpha}{t_\beta}}{1+\frac{t_\alpha}{t_\beta}}
= \frac{1+\frac{1}{t_\beta^2} \frac{k_D}{k_U}}{1-\frac{1}{t_\beta^2} \frac{k_D}{k_U}}
.
\ee
For $|k_U/k_D| \sim 1$,
if $t_\beta$ is larger than about 3 (say),
then
\be
\sin{(\beta - \alpha)}
\sim
\sin{(\beta + \alpha)}
\left[
1 + \frac{2}{t_\beta^2} \frac{k_D}{k_U}
\right].
\ee
Thus, the second factor in Eq.~\eqref{factors} is very closely
given by $\sin{(\beta + \alpha)}$ already for moderate values of $t_\beta$.
In conclusion,
an experimental constraint of $\mu_{VV} \sim 1$ has a solution
$\sin{(\beta-\alpha)} \sim 1$ for all values of $t_\beta$,
and it also has a solution $\sin{(\beta+\alpha)} \sim 1$ for values of
$t_\beta \gtrsim 3 $.
As an illustration,
we show in Fig.~\ref{fig:1} the constraints
on the  $\sin{\alpha}$-$\tan{\beta}$ plane of a 20\% precision measurement
of $\mu_{VV}$ around the SM value 1.
%
%
The left branch corresponds to the right-sign and lies very close
to the line $\sin{(\beta-\alpha)}=1$ ($k_D=1$), while the right branch
corresponds to the wrong-sign and lies very close
to the line $\sin{(\beta+\alpha)}=1$ ($k_D=-1$).

We note that, because both factors in Eq.~\eqref{factors} get closer to
one in the right-sign and wrong-sign limits,
a moderate precision in $\mu_{VV}$ implies a very precise line
in the $\sin{\alpha}$-$\tan{\beta}$ plane \cite{Fontes:2014tga}.
As shown in detail in section IIB of \cite{Fontes:2014tga},
for $\tan{\beta} = 10$ and a precision of $20\%$ in $\mu_{VV}$,
$\sin^2{(\beta-\alpha)}$ is determined to better than $0.5\%$
in the wrong-sign branch.

\section{\label{sec:decay}The $h \rightarrow \Upsilon \gamma$ decay in 2HDM}

\subsection{Decay rate}

\begin{figure}[h!]
\centering
\includegraphics[width=0.65\linewidth]{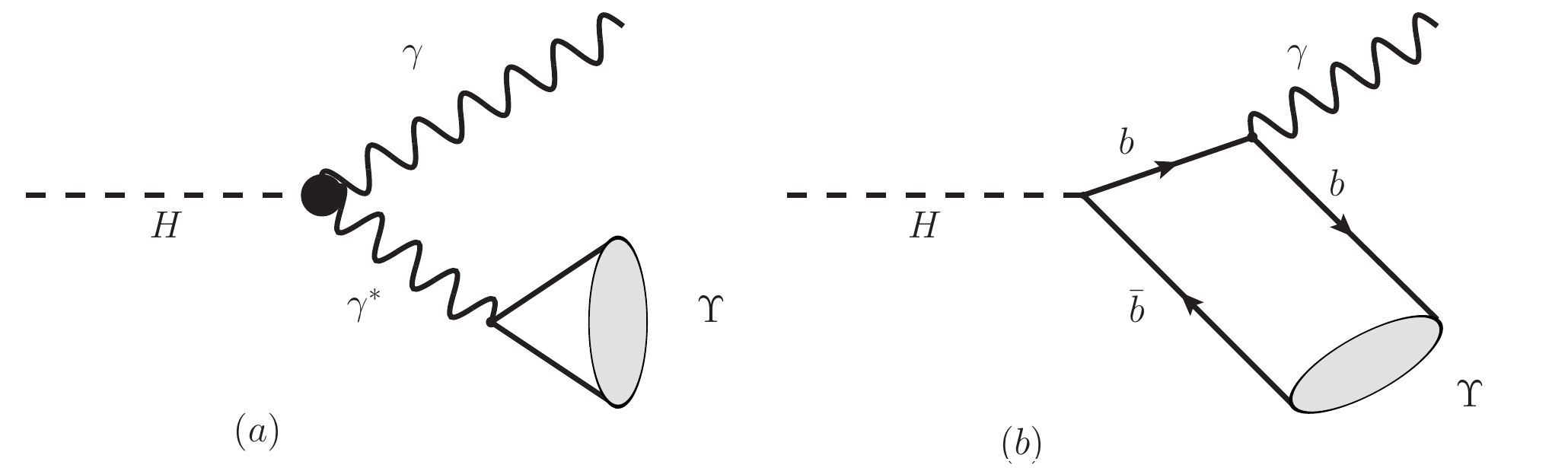}
\caption{Feynman diagrams contributing to the $h \to \Upsilon \gamma$ process.
The diagrams originate from two different couplings:
(a) loop induced $h\gamma\gamma$ (indirect) coupling;
(b)  $h b \bar{b}$ Yukawa (direct) coupling.}
\label{fig:2}
\end{figure}

The $h \rightarrow \Upsilon \gamma$ decay rate may be written as
\be
\Gamma[h \rightarrow \Upsilon \gamma]
=\frac{1}{8 \pi} \frac{m_h^2 - m_\Upsilon^2}{m_h^2}
\left| {\cal A}_\textrm{direct} + {\cal A}_\textrm{indirect} \right|^2 .
\ee
The direct diagram is shown in Fig.~\ref{fig:2}(b) and arises from
the direct $h b \bar{b}$ coupling ($k_D$).
The indirect diagram is shown in Fig.~\ref{fig:2}(a) and arises from the
effective $h \gamma \gamma$ with a virtual photon morphing
into an $\Upsilon$.
%
%

We adapt the calculations of Ref.~\cite{Bodwin:2013gca} to the 2HDM,
and write
\ba
{\cal A}_\textrm{direct}
&=&
-  \eta\, \frac{2}{\sqrt{3}} e\, k_D
\left( \sqrt{2} G_F \frac{m_\Upsilon}{m_h}\right)^{1/2}
\frac{m_h^2- m_\Upsilon^2}{m_h^2- m_\Upsilon^2/2 - 2 m_b^2}\
\phi_0(\Upsilon),
\nonumber\\*[2mm\,]
{\cal A}_\textrm{indirect}
&=&
\frac{e\, g_{\Upsilon \gamma}}{m_\Upsilon^2}
\left( \sqrt{2} G_F \right)^{1/2} \frac{\alpha}{\pi}
\frac{m_h^2- m_\Upsilon^2}{\sqrt{m_h}} \frac{X}{4},
\label{amps}
\ea
where $G_F$ is Fermi's constant, $e$ is the positron charge, $k_D$ is given
in Eq.~\eqref{kUkD}, $m_\Upsilon$ and $m_b$ are the
$\Upsilon$ and b-quark masses,
$\alpha$ is the fine-structure constant,
$\phi_0^2(\Upsilon) \sim 0.512$ $\textrm{GeV}^3$ is the wave function of
$\Upsilon$ at the origin,
and
\be
g_{\Upsilon \gamma} =
\frac{2}{\sqrt{3}} \sqrt{m_\Upsilon}\, \phi_0(\Upsilon),
\ee
whose magnitude can be determined from
\be
\Gamma[\Upsilon \rightarrow \ell^+ \ell^-] =
\frac{4 \pi \alpha^2(m_\Upsilon)}{3 m_\Upsilon^3} g_{\Upsilon \gamma}^2.
\ee
Our expressions in Eqs.~\eqref{amps} bear three
differences with respect to Eqs.~(14a)-(14b) of Ref.~\cite{Bodwin:2013gca}.
First,
we have included explicitly in ${\cal A}_\textrm{direct}$
the factor $\eta = 0.689$
mentioned at the end of section~IIA of \cite{Bodwin:2013gca},
due to the full NLO corrections \cite{Bodwin:2013gca, Bodwin:2014bpa}.
Second,
we have corrected in ${\cal A}_\textrm{indirect}$ a
$\sqrt{2}$ misprint\footnote{We are grateful
to G. Bodwin for clarifications on this point.
We agree with their Eq.~(12), but have a $\sqrt{2}$ difference with respect
to their Eq.~(14b).}.
Finally,
we have defined ${\cal I} = -X/4$, where $X$ is
the function arising from the calculation of the
effective $h \gamma \gamma$ coupling at one-loop in the 2HDM,
which can be found in appendix B of Ref.~\cite{Fontes:2014xva}.

As shown in \cite{Bodwin:2013gca},
the direct and indirect contributions interfere destructively
in an almost complete manner in the SM,
and $h \rightarrow \Upsilon \gamma$ cannot be detected.
This is also the case in the right-sign solution of the 2HDM.
In contrast, the wrong-sign solution has a constructive interference,
raising the prospects for detection.
This is what we turn to next.

\subsection{The importance of $h \rightarrow \Upsilon \gamma$
for the wrong-sign scenario}

As mentioned,
the experimental measurement of
$\mu_{VV}$ means that the $h VV$ and $h t \bar{t}$ couplings
lie close to their SM values. As a result,
$h \rightarrow \gamma \gamma$ in the 2HDM is still dominated
by the $W$ loop, with a small destructive interference correction
from the top loop.
There are two novelties in the 2HDM.
First,
the alteration of $k_D$.
The bottom loop contribution is negligible in the SM.
It can indeed change sign in the 2HDM,
but, since $\mu_{VV}$ places $|k_D| \sim 1$,
it cannot have a strong impact.
Second,
there is a charged Higgs loop.
This decouples with the mass of the charged Higgs,
but it can still give a contribution of up to ten percent for
values of the charged Higgs mass around 600 GeV.
Such effects are inevitable in the wrong-sign scenario
\cite{Ferreira:2014naa}.
One concludes that only precise measurements of
the $h \rightarrow \gamma \gamma$ decays can yield a
signal for the wrong-sign solution of the 2HDM \cite{Ferreira:2014naa, Fontes:2014tga};
the only method presented thus far.

Here we advocate that $h \rightarrow \Upsilon \gamma$ is a good
candidate to determine the sign of $k_D$.
This occurs precisely because the cancellation is almost complete
in the SM. A change in the sign of $k_D$ means that the interference
becomes constructive, thus increasing by orders of magnitude
the $h \rightarrow \Upsilon \gamma$ decay rate.
This can be used to constrain the wrong-sign solution in the 2HDM.

We have performed a full simulation of the real 2HDM,
including theoretical constraints from bounded
from below potential \cite{Deshpande:1977rw},
perturbative unitarity
\cite{Kanemura:1993hm, Akeroyd:2000wc, Ginzburg:2003fe},
oblique radiative parameters
\cite{Peskin:1991sw, Grimus:2008nb, Baak:2012kk},
and we keep $m_{H^\pm} > 480\textrm{GeV}$ to respect
B-physics constraints.
We include all production mechanisms \cite{Spira:1995mt, Harlander:2012pb, vbf}
and take $\mu_{VV}$, $\mu_{\gamma \gamma}$,
and $\mu_{\tau \tau}$ to lie within 20\% of the SM,
in close accordance with the latest LHC constraints
\cite{Khachatryan:2016vau}.

\begin{figure}[h!]
	\centering
	\includegraphics[width=0.4\linewidth]{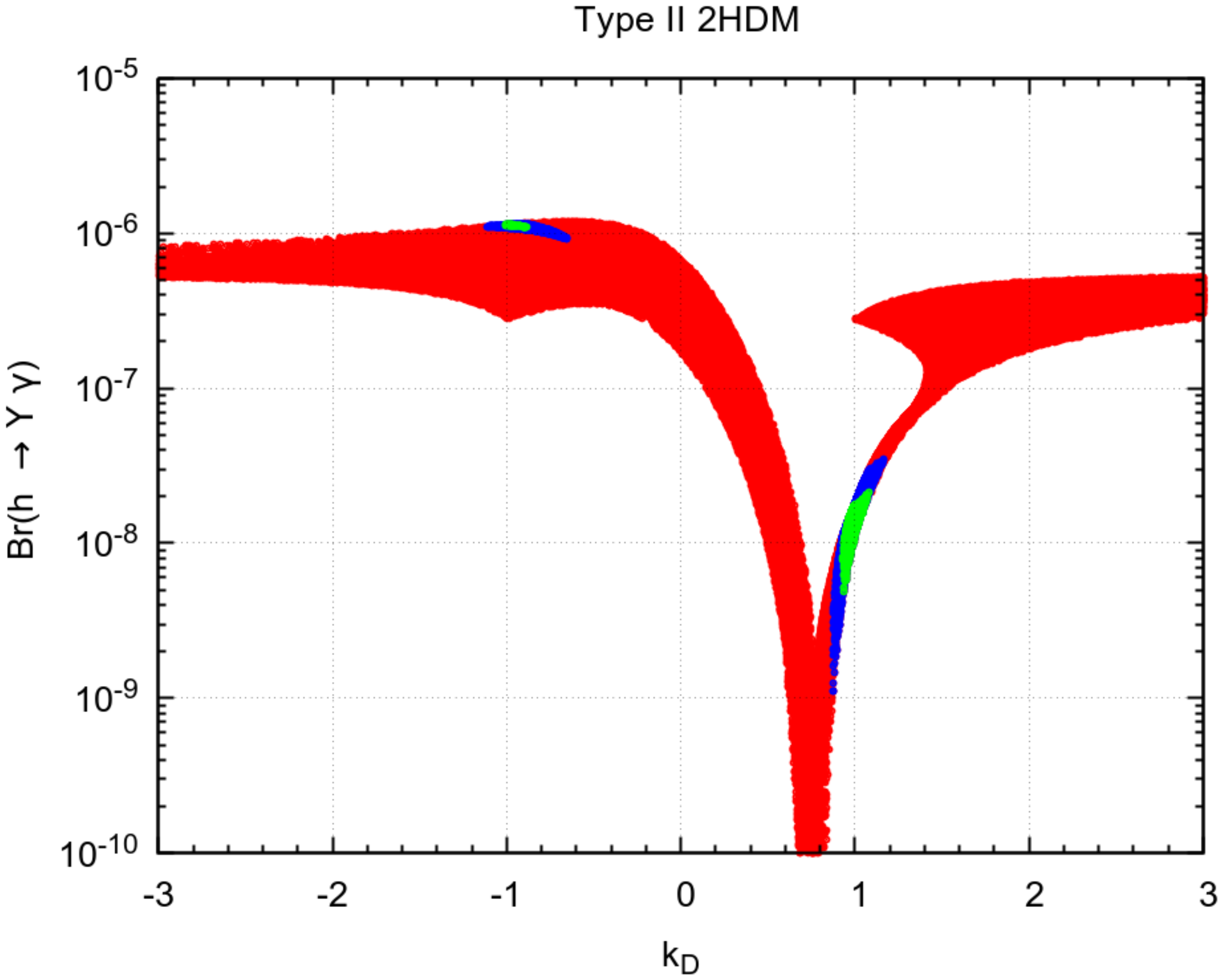}
	\hspace{0.02\linewidth}
	\includegraphics[width=0.4\linewidth]{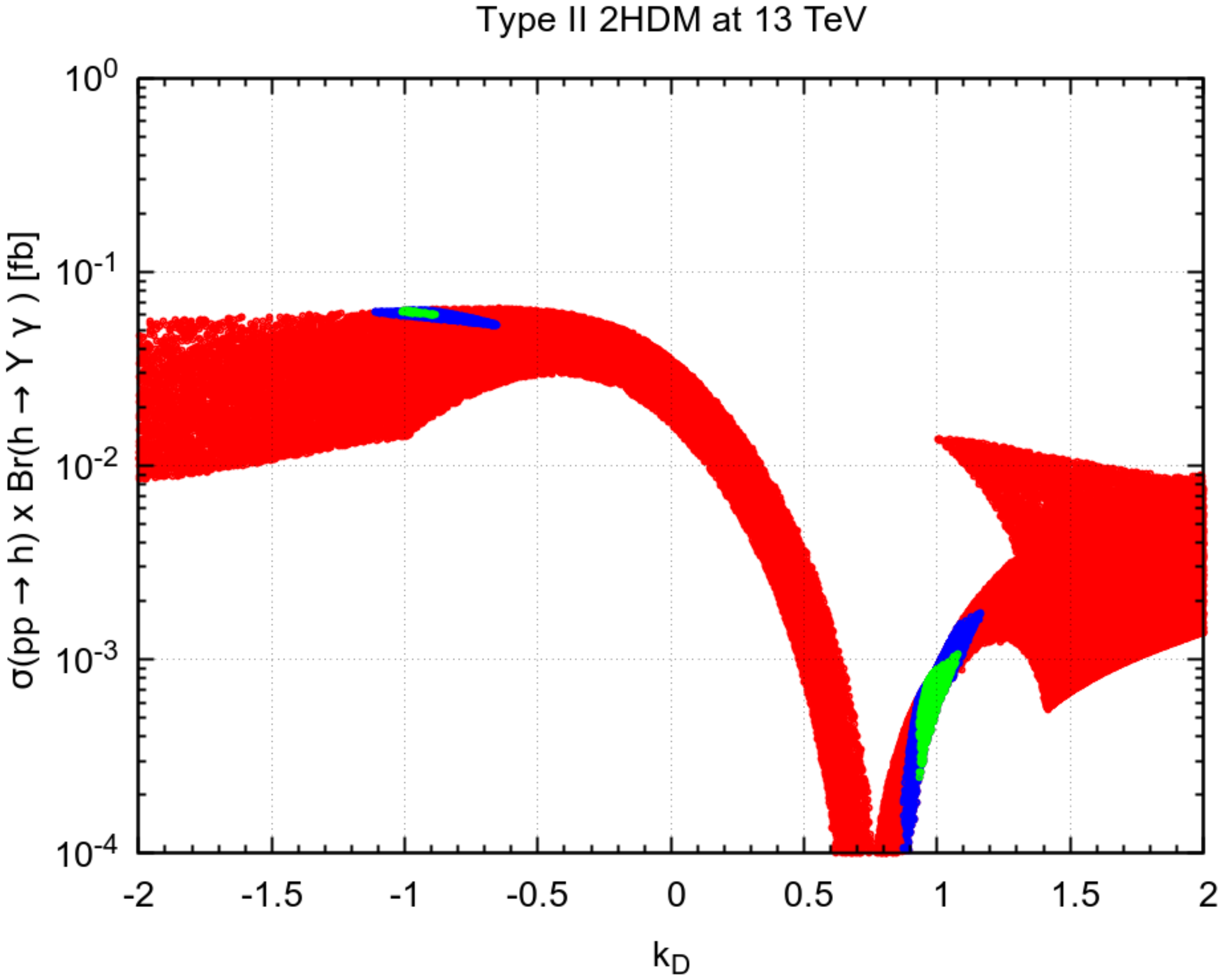}
	\caption{Figure (a): BR($h \rightarrow \Upsilon \gamma$) as a function of $k_D$.
    The red/dark-grey points pass all theoretical constraints in the type II 2HDM.
    The blue/black (green/light-grey) points pass both the theoretical
    constraints and the experimental constraints on
    $\mu_{VV}$, $\mu_{\gamma \gamma }$, and $\mu_{\tau \tau}$ at 20\% (10\%).
    Figure (b): Same plot, but for
    $\sigma(pp \rightarrow h) \times \textrm{BR}(h \rightarrow \Upsilon \gamma)$
    at 13 TeV.
    }
	\label{fig:3}
\end{figure}

The results of our simulation in the type II model are shown in
Fig.~\ref{fig:3}.
%
%
The red/dark-grey points pass all theoretical constraints.
The blue/black (green/light-grey) points pass those and also
$\mu_{VV}$, $\mu_{\gamma \gamma }$, and $\mu_{\tau \tau}$ at 20\% (10\%).
The situation for the flipped model is very similar,
with only very slight differences in the allowed regions,
due to the different dependence on $\mu_{\tau \tau}$.

There are several features of note.
After theoretical constraints,
the simulation allows for a very large range of
$k_D$.
Contrary to what one might naively expect,
having a large $k_D$ does not improve much
the $h \rightarrow \Upsilon \gamma$ branching ratio.
The point is that,
although a large $k_D$ does indeed
increase the direct amplitude,
in accordance with Eq.~\eqref{amps},
in the 2HDM the width of $h$ is dominated by
$h \rightarrow b \bar{b}$,
which also increases with $k_D$.
Once one introduces the experimental constraints,
the values for $k_D$ get restricted to
right-sign ($k_D \sim 1$) and wrong-sign ($k_D \sim -1$)
regions.
As explained in Sec.~\ref{subse:muVV},
this is mostly due to $\mu_{VV}$ and simple
trigonometry \cite{Fontes:2014tga}.
Finally,
one sees that,
due to the same destructive interference at play in the SM,
the right-sign solution leads to a minute
$h \rightarrow \Upsilon \gamma$ branching ratio
around $10^{-8}$.
In contrast,
the wrong-sign solution leads to constructive
interference and a
$h \rightarrow \Upsilon \gamma$ branching ratio
larger by two orders of magnitude.

The possible experimental reach is best seen in
Fig.~\ref{fig:3}(b),
where we show a simulation of
$\sigma(pp \rightarrow h) \times \textrm{BR}(h \rightarrow \Upsilon \gamma)$
at 13 TeV.
For the wrong-sign, we find a value around $0.06$ fb.
The current run II data lies around 15 fb$^{-1}$ total integrated
luminosity \cite{lumi} and will ultimately achieve around
100 fb$^{-1}$, meaning that a measurement is becoming
possible.
This $0.06$ fb estimate arises from the precise values
taken for $g_\Upsilon$ and the scale chosen for
$\alpha$ in the various steps of the calculation.
A detailed discussion,
including relativistic corrections,
can be found in \cite{Bodwin:2014bpa}.
Our result presents a lower limit on the number
of events,
meaning that detection prospects are likely
to be superior.
Of course,
an even better determination is possible at the
High-Luminosity LHC, allowing for the detection or
completely ruling out of the wrong-sign solution.
We have made a simulation at 14 TeV and obtain the expected
increase of about $15\%$ from $0.06$ fb into
around $0.07$ fb, in both Type II and Flipped.

\section{\label{sec:conclusions}Conclusions}

The decay $h \rightarrow \Upsilon \gamma$ is very small in the SM,
due to a cancellation between the direct and indirect diagrams.
In contrast, in theories with a negative $hbb$ coupling,
the interference becomes constructive and the rate is increased
by orders of magnitude.
We have studied this effect on the wrong-sign solution
of the Type II and flipped 2HDM.
We make detailed predictions for the number of events
consistent with current bounds on the 2HDM and
prove that searches for $h \rightarrow \Upsilon \gamma$
constitute a viable and clean method to constrain
the wrong-sign solution, especially at a high luminosity facility.

\vspace{1ex}

\begin{acknowledgments}
We are grateful to G. Bodwin for discussions and to R. Santos
for carefully reading the manuscript.
This work is supported in part by the Portuguese
\textit{Funda\c{c}\~{a}o para a Ci\^{e}ncia e Tecnologia} (FCT)
under contract UID/FIS/00777/2013.
\end{acknowledgments}

\end{document}